\documentclass[twocolumn,showpacs,preprintnumbers,amsmath,amssymb,prb,superscriptaddress]{revtex4}

\usepackage{color}
\usepackage{graphicx}
\usepackage{commath}
\usepackage {tikz}
\usepackage{pgfplots}
\usepackage{feynmp}

\pgfplotscreateplotcyclelist{farbverlauf1}{%
  {solid,black},
  {dashed,blue},
  {densely dotted,red},
  {densely dashed,green},
}
\pgfplotscreateplotcyclelist{farbverlauf2}{%
  {solid,black},
  {dashed,blue},
  {densely dotted,blue!50!red},
  {densely dashed,red},
  {loosely dotted,red!50!green},
  {loosely dashed,green},
  {dotted, green!50!black},
}

\makeatletter
\def\endfmffile{%
  \fmfcmd{\p@rcent\space the end.^^J%
          end.^^J%
          endinput;}%
  \if@fmfio
    \immediate\closeout\@outfmf
  \fi
  \ifnum\pdfshellescape=\@ne
    \immediate\write18{mpost \thefmffile}%
  \fi}
\makeatother

\begin{document}

\title{
Finite frequency noise properties of the non-equilibrium Anderson impurity model}

\author{Christoph P. Orth}
\affiliation{Institut f\"ur Theoretische Physik, Universit\"at Heidelberg,
D--69120 Heidelberg, Germany}
\affiliation{Departement Physik, Universit\"at Basel, CH--4056 Basel, Switzerland}
\author{Daniel F. Urban}
\affiliation{Physikalisches Institut, Albert--Ludwigs--Universit\"at Freiburg,
D--79104 Freiburg, Germany}
\author{Andreas Komnik}
\affiliation{Institut f\"ur Theoretische Physik, Universit\"at Heidelberg,
D--69120 Heidelberg, Germany}

\date{\today}

\begin{abstract}
We analyze the spectrum of the electric-current autocorrelation function (noise power) in the Anderson impurity model biased by a finite transport voltage. Special emphasis
is placed on the interplay of non-equilibrium effects and electron-electron interactions. Analytic results are presented for a perturbation expansion in the interaction strength $U$.
Compared to the non-interacting setup we find a suppression of noise for finite frequencies in equilibrium and an amplification in non-equilibrium.
Furthermore, we use a diagrammatic resummation scheme to obtain non-perturbative results in the regime of intermediate $U$. At finite voltage, the noise spectrum shows sharp
peaks at positions related to the Kondo temperature instead of the voltage.
\end{abstract}


\maketitle

\section{Introduction}

The Anderson impurity model (AIM) is probably the best studied interacting quantum impurity system.\cite{Anderson1961,Hewson1997} Nonetheless, there are still a number of open
questions, especially concerning the physical properties of its non-equilibrium multi-terminal incarnations. These are archetypical models to describe transport
properties of nanoscale quantum dots. In recent years not only the classical transport characteristics, such as linear conductance and full current-voltage relations, have been studied, but
also more complex quantities, like the full counting statistics, have been explored. Thereby a large variety of different analytical as well as numerical techniques have been applied to the AIM (see for
instance~[\onlinecite{Hershfield1992,Hershfield1993,Gull2011,Anders2010,Eckel2010,Muhlbacher2011,Gogolin2006,PhysRevB.76.085342,Oguri2001a,PhysRevB.84.235314,PhysRevB.79.235336,PhysRevLett.102.146803,PhysRevLett.87.156802,PhysRevLett.87.236801,PhysRevLett.84.3686,PhysRevLett.76.487}]). However, most of the studies have concentrated on static
or integral energy independent properties. Aiming at the ultimate practical applications of the devices in question, the knowledge of their frequency-dependent noise
would undoubtedly be of great advantage. Currently, research in that direction appears to gain momentum.\cite{Rothstein2009,Moca2011,Gabdank2011} The frequency-dependent noise
is not of mere theoretical interest as even the frequency-dependent current cumulants of higher order are by now measurable in experiments.\cite{Ubbelohde2012}

Depending on its parameters the AIM shows many interesting phenomena, mostly with signatures in transport properties of the two-terminal setup. Probably the
most prominent one is the Kondo effect which is generally only visible for large on-site repulsion $U$. Very few analytical methods starting from the proper AIM are
able to account for this correlation effect, as the analytical solution for the AIM at arbitrary $U$ in non-equilibrium does not exist. As far as the integral
transport characteristics are concerned, it was shown in Refs.\ [\onlinecite{Oguri2001a,Oguri2002,Gogolin2006}] that the current-voltage relation as well as the full counting
statistics can be calculated up to the cubic order in applied voltage and for arbitrary $U$. For the frequency-dependent noise the non-interacting situation [resonant level
model (RLM)] in non-equilibrium was addressed in Refs.\ [\onlinecite{Rothstein2009},\onlinecite{BlanterButtiker}].
A number of interesting frequency-dependent noise results were obtained by use of the equation-of-motion technique,\cite{Ding1997} and Filippone \emph{et al.} recently investigated similar quantities in a slightly different context.\cite{Filippone2011} Nevertheless, the general crossover from weak to
strong $U$ --- eventually entering the Kondo regime --- was not yet addressed for the frequency-dependent noise. With this work we want to close this gap and provide results valid for arbitrary bias voltage and hybridization parameter.

The paper is organized as follows. In Sec.\ \ref{pert} we start with a simple perturbative expansion in $U$. Using the recently derived analytical results for the non-equilibrium self-energy
we analyze the interplay of finite voltage and on-site repulsion and identify the related features in the noise spectrum. In Sec.\ \ref{resum} we derive a diagrammatic resummation
scheme (which is similar to the one used in Ref.\ [\onlinecite{Hershfield1992a}] for the shot noise) that enables us to generate non-perturbative results. We find
that for larger $U$ the noise spectra show distinct features which can be related to the Kondo effect. Section \ref{discussion} concludes the paper with a summary of the results.

\section{Perturbative expansion in $U$}
\label{pert}

We model the system by the two-terminal AIM Hamiltonian
\begin{eqnarray}
  H &=& H_L + H_R + H_d + H_T + H_I \\
  H_d &=& \sum_{\sigma} \Delta \, d^\dagger_\sigma d_{\sigma} \nonumber \\
  H_T &=& \sum_{\sigma} \gamma (d^\dagger_\sigma L_\sigma + d^\dagger_\sigma R_\sigma + \mathrm{H.c.}) \nonumber \\
  H_I &=& U d^\dag_\uparrow (t) d_\uparrow (t) d^\dag_\downarrow (t) d_\downarrow (t) \nonumber \, .
\end{eqnarray}
$H_d$ describes the spin($\sigma$) degenerate dot level at energy $\Delta$ with the fermionic creation and annihilation operators $d^\dag_\sigma$ and $d_\sigma$.
$H_T$ is the tunneling Hamiltonian with tunneling amplitude $\gamma$ to the left ($L$) and right ($R$) leads\footnote{The symmetry of the coupling to the electrodes is not essential and the whole analysis can as well be performed for the asymmetric case $\gamma_L \neq \gamma_R$.}
and $H_I$ models Coulomb interaction on the dot. $H_L$ ($H_R$) describe free electrons in the left (right) lead, held at constant chemical potential $\mu_L=V/2$ ($\mu_R=-V/2$).
The principal quantity to calculate is the current noise spectrum
\begin{eqnarray} \label{eq:NoiseDef}
  S(\omega)=\int \dif t \, e^{- i \omega t} \left( \langle I(t) I(0) \rangle - \langle I \rangle^2 \right),
\end{eqnarray}
given as the Fourier transform of the current autocorrelation function.
As was pointed out in Refs.~[\onlinecite{Lesovik1997,Gavish2000,Billangeon2009}] this unsymmetrized definition
of $S(\omega)$ is the most natural one for quantum systems:
negative frequencies describe the spectrum of fluctuations which emit energy to the environment, positive frequencies are related to the absorption spectrum. Furthermore, in order
to be as close to the experimental situation as possible, we define the total current operator through the constriction as the average through the two contacts
(we use units in which $e=\hbar=c=1$ throughout),
\begin{multline} \label{eq:AIMcurrent}
  I(t) 
  = i \frac{\gamma}{2} \sum_{\sigma=\uparrow, \downarrow} \left[ L^\dagger_\sigma(t)  d_\sigma(t) - R^\dagger_\sigma(t)  d_\sigma(t)-\mathrm{H.c.} \right] \, .
\end{multline}
As a first step, we evaluate the current correlations $\langle I(t) \, I(t') \rangle$ in the non-interacting case ($U=0$), when the spin sector completely factorizes.
The noise can be rewritten in terms of
Keldysh Green's functions (GF) $G^{kl}_{0,\alpha\beta}(t-t')=-i \langle T_C \, \alpha(t) \beta^\dag(t') \rangle_0$ where $\alpha, \beta$ are lead ($L$,$R$) or dot ($d$)
creation or annihilation operators. $k$ and $l$ are Keldysh indices of the times $t$ and $t'$ and $T_C$ is the Keldysh time ordering operator. The
average $\langle\;\cdot\;\rangle_0$ is over the non-interacting part of the system only for which the GF are known exactly. In particular, we define the dot GF by $D_0^{kl}=G^{kl}_{0,dd}$
for which one obtains\cite{Caroli1971,Hershfield1992,Gogolin2006}
\begin{equation} \label{eq:dotGF}
 \boldsymbol D_0(\omega) = \left( \begin{matrix} \frac{\omega-\Delta + 2 i \Gamma \left[n(\omega) -1 \right]}{(\omega-\Delta)^2+4\Gamma^2} & \frac{2 i \Gamma n(\omega)}{(\omega-\Delta)^2+4\Gamma^2} \\ \frac{2 i \Gamma \left[n(\omega) -2 \right]}{(\omega-\Delta)^2+4\Gamma^2} & \frac{-\omega+\Delta + 2 i \Gamma \left[n(\omega) -1 \right]}{(\omega-\Delta)^2+4\Gamma^2} \end{matrix} \right).
\end{equation}
Here we defined $n(\omega)=\theta(V/2-\omega)+\theta(-V/2-\omega)$ with the unit-step function $\theta$.
The electrodes are described in the framework of the wide flat-band model with constant density of states  $\rho_0$. Furthermore, we use $2\Gamma=2\pi \gamma^2 \rho_0$ as
the unit of energy if $\Gamma$ does not occur explicitly. In the following we restrict the analysis to the zero-temperature case as one expects only a smear-out effect
in the noise spectrum at finite temperature.
Within these approximations we obtain the following analytical result for $U=0$,
\begin{multline} \label{eq:AIM0ONP1}
  S(\omega) =\frac{1}{2 \pi} \sum_{\sigma,\sigma'=\pm} \Biggl\{ \frac{\sigma}{2} \, Q(\omega,\sigma V) \times \\
	\times \Bigl[\arctan\left(V/2 +\sigma' \Delta +\sigma \omega\right)+\arctan\left(V/2 +\sigma' \Delta\right)\Bigr]  \\
	+\frac{1}{\omega} \, P(\omega,\sigma V) \, \ln\left[\frac{(V/2 +\sigma' \Delta +\sigma \omega)^2+1}{(V/2 +\sigma' \Delta)^2+1}\right] \Biggr\}.
\end{multline}
Here $P(x, y) =\theta(x)-\theta(y+x)$ is a `window' function and $Q(x, y) =\theta(x)+\theta(y+x)$ is a double-step function.
Our result (\ref{eq:AIM0ONP1}) is in agreement with the numerical results of Ref. [\onlinecite{Rothstein2009}].
It can be derived from the general expression for a two-terminal system with the Lorentzian-shaped energy dependent transmission coefficient
(see e.~g. [\onlinecite{BlanterButtiker}])
\[
T(\omega) = 4\Gamma^2/((\omega-\Delta)^2 + 4\Gamma^2) \, .
\]
The generic features of the non-interacting noise spectra are characteristic steps in the unsymmetrized noise spectrum (and cusps in the symmetrized version) at energies
corresponding to $\pm\Delta \pm V/2$.

\begin{figure}[!ht]
  \centering
  \includegraphics{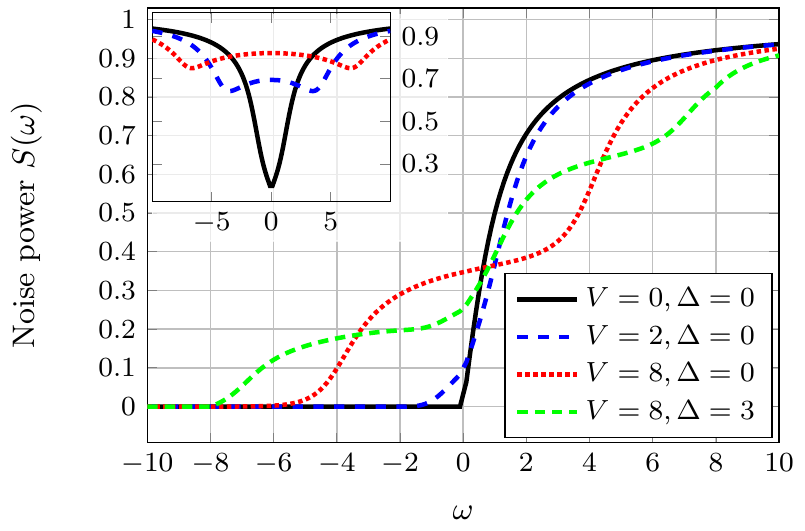}
  \caption{Unsymmetrized noise spectrum for the non-interacting case. The negative/positive $\omega$ part can be interpreted as
    emission/absorption of energy by the system to/from the environment. Steps emerge whenever the energy $\omega$ of a fluctuation is about
    $\pm V/2\pm\Delta$. The inset shows the symmetrized spectrum for the same parameters.}
  \label{fg:NoIntNoise}
\end{figure}

It is well known that the most prominent feature of the Kondo effect is the emergence of a sharp resonance around the Fermi edge of the metallic continua coupled to an 
impurity. An effective model at energies much smaller than the Kondo scale $T_K$ is a resonant level model, where the role of hybridization $\Gamma$ is given by 
$T_K$\cite{Nozieres1974,Hewson1997}. Therefore our result (\ref{eq:AIM0ONP1}) can formally be used to model the behaviour of the noise in the deep Kondo limit. Indeed, an 
explicit fit of the data from [\onlinecite{Zarchin2008}] shows a very good agreement with the experiment provided that effective  $\Gamma \approx 4.9 \mu$eV.

Now we turn to the perturbative expansion in $U$. As long as one is interested in the electron-hole symmetric case $\Delta = - U/2$ there is no linear contribution
in $U$ and the expansion starts with terms of quadratic order.\footnote{Interestingly, this is true even for the correlation of currents with anti-parallel
spins.\cite{Diplomarbeit_C.Orth}} We restrict the following analysis to this special case since it contains most of the interesting features.
In contrast to the interaction-free case there is no spin-sector factorization anymore. The expression for the noise can still be obtained in a straightforward way. It contains a large number of terms which can be subdivided into two large classes.
We discuss exemplarily at full extent one representative term which contains dot operators and those for the left electrode. All other terms can be treated equivalently.
The two classes of terms are the \emph{parallel-spin} contribution
$S^{\uparrow \uparrow}_{d L \, dL} = \langle d^\dag_\uparrow (t ) L_\uparrow (t) d^\dag_\uparrow (0) L_\uparrow(0)\rangle$ and the \emph{anti-parallel-spin} contribution
$S^{\uparrow \downarrow}_{d L \, dL} =\langle d^\dag_\uparrow (t ) L_\uparrow (t) d^\dag_\downarrow (0) L_\downarrow(0)\rangle$.
In comparison, both have very different structures and thus need to be considered separately.

\begin{figure}[ht]
\begin{minipage}[c]{0.32\linewidth}
  \textbf{(1)} \vspace{.2cm} \hspace{1cm}

  \centering
  \includegraphics{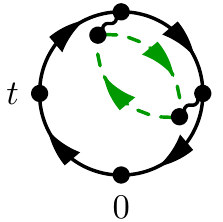}
\end{minipage}
\begin{minipage}[c]{0.32\linewidth}
  \textbf{(2)} \vspace{.2cm} \hspace{1cm}

  \centering
  \includegraphics{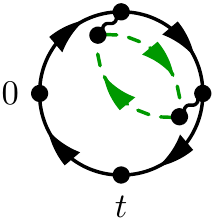}
\end{minipage}
\begin{minipage}[c]{0.32\linewidth}
  \textbf{(3)} \vspace{.2cm} \hspace{1cm}

  \centering
  \includegraphics{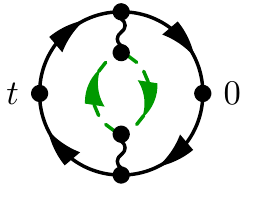}
\end{minipage}
\begin{minipage}[c]{\linewidth}
 \vspace{.7cm} \textbf{(4)} \hspace{2.5cm}

  \centering
  \includegraphics{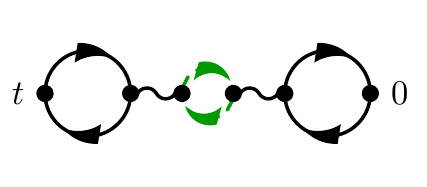}
\end{minipage}
\caption{Diagrammatic representation of the parallel-spin current correlations in the second order perturbation theory for the electron-hole symmetric case. All arrows represent 
  non-equilibrium Keldysh GF exact in the $U=0$ model. The wiggly line represents instantaneous Coulomb interaction. Dashed green arrows describe particles with opposite spin
  than solid black ones. $t$ and $0$ label the two external points of the currents in equation \ref{eq:NoiseDef}. Information about Keldysh indices and lead affiliation
  (especially if of homogeneous or inhomogeneous kind) of the GF are omitted.}
\label{fg:FeynPertEq}
\end{figure}

For the parallel-spin contribution in second order we obtain
\begin{multline} \label{eq:NPequalFreqInt}
  S_{dLdL}^{\uparrow\uparrow}(\omega) = - \frac{\gamma^2}{4} \int \! \frac{\dif{\omega_1}}{2 \pi} \, G_{0,Ld}^{+-}(\omega_1) G_{0,Ld}^{-+}(\omega_1-\omega) \\
  - U^2\frac{\gamma^2}{4} \sum_{k,l=\pm} k l \int \! \frac{\dif{\omega_1}\dif{\omega_2}\dif{\omega_3}}{(2 \pi)^3} \left( B_1 + B_2 + B_3 - B_4 \right) \, ,
\end{multline}
with the definitions
\begin{align}
  B_1 = &D_0^{lk}(\omega_1) D_0^{kl}(\omega_1 + \omega_2 - \omega_3) D_0^{lk}(\omega_2) D_0^{k-}(\omega_3) \nonumber \\
	&\times G_{0,Ld}^{-+}(\omega_3-\omega) G_{0,Ld}^{+l}(\omega_3)\, , \nonumber \\
  B_2 = &D_0^{lk}(\omega_1) D_0^{kl}(\omega_1 + \omega_2 - \omega_3) D_0^{lk}(\omega_2) D_0^{k+}(\omega_3) \nonumber \\
	&\times G_{0,Ld}^{+-}(\omega_3+\omega) G_{0,Ld}^{-l}(\omega_3)\, , \nonumber \\
  B_3 = &D_0^{lk}(\omega_1) D_0^{kl}(\omega_1 + \omega_2 - \omega_3) D_0^{l-}(\omega_2) D_0^{k+}(\omega_3-\omega) \nonumber \\
	&\times G_{0,Ld}^{+l}(\omega_3) G_{0,Ld}^{-k}(\omega_2-\omega)\, , \nonumber \\
  B_4 = &D_0^{lk}(\omega_2) D_0^{kl}(\omega_2 - \omega) D_0^{l+}(\omega_3-\omega) D_0^{k-}(\omega_1) \nonumber \\
	&\times G_{0,Ld}^{+l}(\omega_3) G_{0,Ld}^{-k}(\omega_1-\omega) \nonumber \,  .
\end{align}
The corresponding diagrammatic representation is shown in Fig.~\ref{fg:FeynPertEq}.
While the diagrams (1) and (2) can conveniently be reduced to the self-energy of the second order, for which analytic expressions exist even in
non-equilibrium,\cite{Muhlbacher2011} the evaluation of the other two diagrams involves an extra energy integration, which can conveniently be performed numerically.

\begin{figure}[ht]
\begin{minipage}[c]{0.49\linewidth}
  \textbf{(1)} \vspace{.1cm} \hspace{1.2cm}

  \centering
  \includegraphics{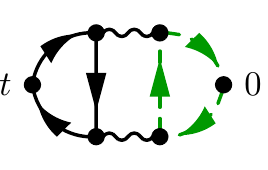}
\end{minipage}
\begin{minipage}[c]{0.49\linewidth}
  \textbf{(2)} \vspace{.1cm} \hspace{1.2cm}

  \centering
  \includegraphics{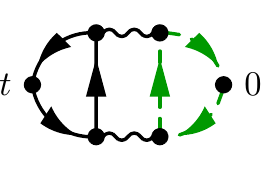}
\end{minipage}
\caption{Diagrammatic representation of the anti-parallel-spin current correlations in the second order perturbation theory for the electron-hole symmetric case.}
\label{fg:FeynPertOpp}
\end{figure}

No reduction in terms of the self-energy is possible for the anti-parallel-spin configuration contributions given by
  \begin{equation} \label{eq:NPunequalFreqInt}
    S_{dLdL}^{\uparrow\downarrow}(\omega) = - U^2\frac{\gamma^2}{4} \sum_{k,l=\pm} k l \int \! \frac{\dif{\omega_1}\dif{\omega_2}\dif{\omega_3}}{(2 \pi)^3} \left( D_2 + D_3 \right) \, .
  \end{equation}
Here
  \begin{align}
    D_2 = &D_0^{lk}(\omega_1) D_0^{kl}(\omega_1 + \omega_2 - \omega_3) D_0^{l-}(\omega_2) D_0^{k+}(\omega_3-\omega) \nonumber \\
	  &\times G_{0,Ld}^{+l}(\omega_3) G_{0,Ld}^{-k}(\omega_2-\omega)\, , \nonumber \\
    D_3 = &D_0^{lk}(\omega_1) D_0^{k-}(\omega_1 + \omega_2 - \omega_3) D_0^{lk}(\omega_2) D_0^{k+}(\omega_3-\omega) \nonumber \\
	  &\times G_{0,Ld}^{-l}(\omega_1+\omega_2-\omega_3) G_{0,Ld}^{+l}(\omega_3) \nonumber \, .
  \end{align}
which can be represented by the diagrams shown in Fig.~\ref{fg:FeynPertOpp}. It turns out that all diagrams contribute to the noise power with the same order of magnitude.

An important benchmark for the full result is the limiting behavior at zero-frequency [the shot noise $S(0)$] for the second order in $U$ correction, for which analytical
results exist for small voltages\cite{Hershfield1992a,Gogolin2006}
\begin{equation}
 S(0) = 6 U^2 |V|^3/\pi^3 \Gamma^4 \, .
\end{equation}
This limit is perfectly reproduced by our calculation of the full noise up to voltages $V/\Gamma\approx1$.
Note that it was shown in Ref.\ [\onlinecite{Rothstein2009}] that $S(\omega)$ is a purely real quantity which is also reflected in our calculation.

\begin{figure}[!ht]
  \centering
  \includegraphics{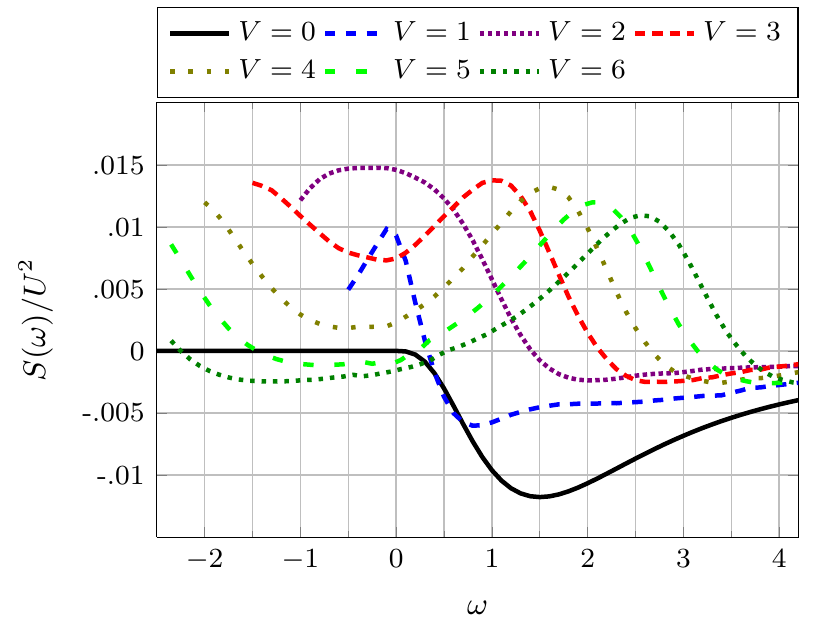}
  \caption{Correction to the noise spectrum in the second order perturbation theory in the resonant configuration $\Delta=-U/2$ in units of $U^2$. }
  \label{fg:PertNoise}
\end{figure}
Figure \ref{fg:PertNoise} shows the second order correction to the full noise as a function of frequency for different values of $V$.
The overall scale of this contribution is about 1\% of the $U=0$ value for $U/\Gamma=2$ and scales with $U^2$. As $U=0$ is a fixed point in the RG-sense, it should be valid for small $U$.
In equilibrium a general suppression of noise is observed (see solid curve in Fig. \ref{fg:PertNoise}). Since the dot can be considered to have a finite capacitance one is
effectively confronted with a high-pass filter, which leads to the above mentioned effect. The situation is completely different in case of a finite bias.
First of all, even a very small $V$ leads to non-vanishing noise around $\omega=0$, where the noise is zero in equilibrium. This remarkable behavior can be traced back to the
imaginary part of the self-energy, which is algebraic in $\omega$ in equilibrium but immediately acquires a constant $\sim V$ term after a bias is switched on. As this
part of the self-energy describes dissipation, it is natural that $V\neq 0$ leads to enhanced fluctuations in the low-energy sector. The form
of this contribution is nearly Lorentzian with a width $\sim \Gamma$, as opposed to the simple step in the non-interacting case. For
$V>\Gamma$ this peak splits into two --- similar to the doubling of the step in the non-interacting case --- and as the voltage grows both peaks shift roughly linearly in $V$. From the mathematical point of view this behavior is understandable
as the noise correction is composed out of GF with Lorentzian shape that are cut off by the chemical potentials in the leads. This step (non-interacting) vs peak (interacting)
behavior of the noise should be clearly distinguishable in the relevant experiments.

\section{Diagrammatic resummation scheme}
\label{resum}

Although the perturbative expansion up to $U^2$ might yield reasonable results\cite{Muhlbacher2011} --- at least for large voltages --- this is not the generic situation
encountered in experiments (see e.~g.\ Ref.\ [\onlinecite{NatNanoFranceschi}]). The same applies to the opposite limit of infinitely strong interactions for which a number
of interesting results became available recently.\cite{Moca2011}. In the following, we address the range of intermediate $U$ by using a diagram resummation technique,
which was presented in a simpler version earlier in Ref.\ [\onlinecite{Hershfield1992a}].

\begin{figure}[ht] \centering
 $S(\omega)=$
 \begin{minipage}[c]{0.22\linewidth}
   \includegraphics{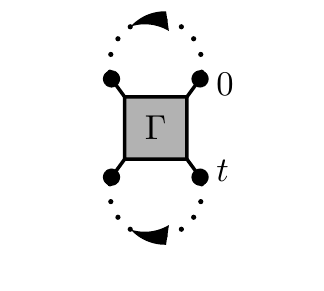}
 \end{minipage}
 \caption{Reduction of the noise to the 2-particle GF $\Gamma$. A distinction between correlations of anti-parallel-spin particles and of parallel-spin particles
    has to be made but was omitted in this figure. To express the noise in this way one needs two free lead GF $g_{L/R}$ expressed by the dotted arrows.}
 \label{fg:FeynIntOutLeads}
\end{figure}

The full noise can be expressed with help of a two-particle GF (which we call $\Gamma^{klmn}$, with 4 Keldysh indices),
\begin{eqnarray}
   \Gamma_{\sigma, \sigma'}^{klmn}(t_1, t_2, t_3, t_4) =& \langle T_C \, d_\sigma(t_1^k) \, d^\dagger_\sigma(t_2^l) \, d_{\sigma'}(t_3^m) \, d^\dagger_{\sigma'}(t_4^n) \rangle\nonumber \\
\end{eqnarray}
which involves only dot operators. Therefore, two free lead GFs are separated from
the average values, see Fig. \ref{fg:FeynIntOutLeads}. Then the frequency-dependent noise can be expressed as
\begin{multline} \label{eq:RPAfullNPFT}
 S(\omega)=-\frac{1}{4} \sum_{k,l=\pm} \; \sum_{\sigma, \sigma' = \uparrow, \downarrow} k l \int \frac{\dif \omega_1 \dif \omega_2}{(2 \pi)^2} \times \\
  \times \Bigl[ \Gamma_{\sigma, \sigma'}^{k+l-}(\omega_1,\omega+\omega_1,\omega_2) g_-^{+k}(-\omega_1)g_-^{-l}(-\omega_2) \\
  +\Gamma_{\sigma, \sigma'}^{+k-l}(\omega_1+\omega,\omega_1, \omega_2) g_-^{k+}(-\omega_1)g_-^{l-}(-\omega-\omega_2) \\
  -\Gamma_{\sigma, \sigma'}^{k+-l}(\omega_1, \omega+\omega_1, \omega_2) g_-^{+k}(-\omega_1)g_-^{l-}(\omega-\omega_2) \\
  -\Gamma_{\sigma, \sigma'}^{+kl-}(\omega_1+\omega,\omega_1, \omega_2) g_-^{k+}(-\omega_1)g_-^{-l}(-\omega_2) \Bigr] \\
  +\frac{1}{4} \sum_{k=\pm=-l} \int \frac{\dif \omega_1}{2 \pi} D_0^{lk}(\omega_1-k\omega)g_+^{kl}(\omega_1) \, .
\end{multline}
Here we have taken the four dimensional Fourier transform of $\Gamma^{klmn}(t_1,t_2,t_3,t_4)$ and neglected its 4$^{\rm th}$ argument, which is possible because of energy conservation.
Furthermore, we have used the definition
\begin{eqnarray}
  g_\pm^{kl}(\omega)=&\gamma^2 \left[ g_L^{kl}(\omega)\pm g_R^{kl}(\omega)\right]
\end{eqnarray}
where the left/right lead free GF are denoted by $g_{L/R}$.

\begin{figure}[ht]
\begin{minipage}[c]{0.12\linewidth}
  \includegraphics{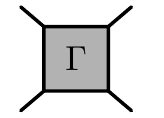}
\end{minipage}
$\; \approx-i U$ \hspace{-.7cm}
\begin{minipage}[c]{0.12\linewidth}
  \includegraphics{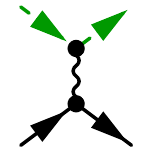}
\end{minipage}
\hspace{-.3cm} $-U^2$ \hspace{-.4cm}
\begin{minipage}[c]{0.12\linewidth}
  \includegraphics{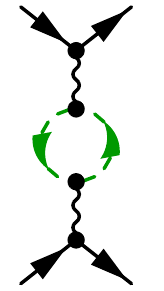}
\end{minipage}
\hspace{.1cm} $+i U^3$ \hspace{-.7cm}
\begin{minipage}[c]{0.12\linewidth}
  \includegraphics{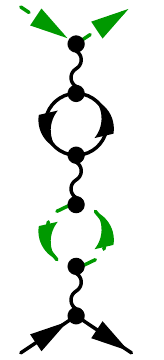}
\end{minipage}
\hspace{-.3cm} $+U^4$ \hspace{-.4cm}
\begin{minipage}[c]{0.12\linewidth}
  \includegraphics{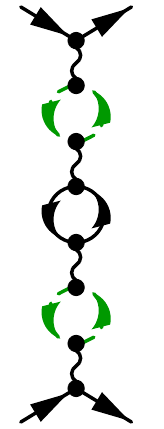}
\end{minipage}
$+\cdots$
  \caption{The complete summation over all diagrams that contribute to $\Gamma$ in our approximation. All arrows in this case represent the dot GF $D_0$ of the $U=0$
    model which is exact in tunneling. Every second diagram is for anti-parallel-spin current correlations.}
  \label{fg:RPAFeynRow}
\end{figure}

In general, the perturbation expansion for this kind of quantity is rather involved. In the following we concentrate on the contribution due to the ladder
diagrams of the type shown in Fig.~\ref{fg:RPAFeynRow} which lead to an effective, screened interaction. These diagrams are connected at both ends as shown
in Fig.~\ref{fg:FeynIntOutLeads} to obtain the noise. The lowest order contribution of this row is also contained in the perturbation theory as Fig.~\ref{fg:FeynPertEq}{\bf (4)}.
The diagrams shown in Fig.~\ref{fg:RPAFeynRow} describe processes in which an electron pair is consecutively scattered on \emph{uncorrelated} electron-hole pairs (bubbles).
It is well-known that the neglected processes are unimportant in the case of bulk extended fermionic systems of high density and the respective approximation is the familiar random phase
approximation (RPA). In the RPA approximation one sums over the diagrams with leading contribution for each order. This argument does not apply in our case as a
zero-dimensional dot the concept of a density does not exist. However, with help of the GF in Eq. (\ref{eq:dotGF}) one can show that more convoluted diagrams should contribute less than the simple
ones we use in this approximation, at least for not too big $\omega$.

Furthermore, diagrams similar to {\bf (1)} and {\bf (2)} from Fig.~\ref{fg:FeynPertEq} are composed of two single particle GF and do not spawn vertex corrections. The frequency-dependent
noise however is an inherent vertex dependent quantity and effects from diagrams of this kind are expected to only result in features which are already well known. To support this
point we calculated the noise also with help of a resummation using the analytical non-equilibrium self-energy \cite{Muhlbacher2011} which revealed no qualitative contribution.
Moreover, we investigated diagrams related to {\bf (3)} from Fig.~\ref{fg:FeynPertEq} by using the screened interaction of Fig.~\ref{fg:RPAFeynRow} in diagram
{\bf (3)}, Fig.~\ref{fg:FeynPertEq}.
 Numerical checks for various parameter constellations confirm that there is no significant contribution from these diagrams.
This strongly suggests that the features we find using our resummation of diagrams similar to
Fig.~\ref{fg:FeynPertEq}{\bf (4)} are not canceled by diagrams which are neglected and are part of the exact solution of the model.


 Although our approximation is very similar in its spirit to the RPA, a precise identification of the small parameter is impossible. Nonetheless, we can roughly estimate the relevant parameter regime by observing that in all diagrams we take into account the loops describe electron-hole production/annihilation processes, in which every electron is annihilated by its `own' hole. Since any recombination process with a different (`alien') hole can only take place by tunneling into the electrode and back the recombination processes with `own' holes are more probable for larger dwelling times $\tau$ of the electrons on the dot. Since $\tau \sim 1/\Gamma$ our approximation is valid when $\Gamma$ is small in comparison to $U$.

On the other hand, Fermi liquid theory for the non-equilibrium transport through an Anderson impurity\cite{Oguri2001a,Oguri2002} implies that the principal observables can be expressed in terms of equilibrium susceptibilities. This is just as in the equilibrium case which was treated in the famous series of contributions by Yamada and Yosida.\cite{Yamada1975a,Yosida1970,Yosida1975}
Interestingly, from the diagrammatic point of view the equilibrium susceptibilities are represented by simple loops. Although it is very difficult to establish a one-to-one correspondence, our approach is a version of a Fermi liquid theory applied to non-equilibrium noise and it is therefore expected to yield adequate results particularly in the low energy sector.

The sum over ladder diagrams shown in Fig. \ref{fg:RPAFeynRow} is equivalent to the solution of a set of 32 linear equations:
\begin{eqnarray}
    \lefteqn{\Gamma_{\uparrow \downarrow}^{klmn}(\omega_1,\omega_2,\omega_3) \approx
    }\quad\nonumber\\
&&
    i U \sum_{p=\pm} p D_0^{k p}(\omega_1) D_0^{p l}(\omega_2) \Gamma_{\uparrow \uparrow}^{p p m n}(\omega_1,\omega_2,\omega_3),
\\
    \lefteqn{\Gamma_{\uparrow \uparrow}^{klmn}(\omega_1, \omega_2, \omega_3) \approx 2 \pi \delta(\omega_3-\omega_2) D_0^{kn}(\omega_1) D_0^{ml}(\omega_3)
    }\quad\nonumber\\
&&
    + i U \sum_{p=\pm} p D_0^{kp}(\omega_1)D_0^{pl}(\omega_2) \Gamma_{\uparrow \downarrow}^{ppmn}(\omega_1,\omega_2, \omega_3).\quad
\end{eqnarray}
An explicit analytical solution is possible but rather lengthy so we discuss the respective numerical results only. From general grounds, $\Gamma^{klmn}$ should satisfy certain
symmetry conditions. Similar to other partial resummation schemes, it vanishes for $\omega_2 \neq \omega_3$. Moreover, it should satisfy
$\Gamma^{klmn}(\omega_1, \omega_2, \omega_2)=\Gamma^{mnkl}(\omega_2, \omega_1, \omega_1)$ due to the symmetry under exchange of the incident particles. Because of
time-reversal symmetry it also suffices the symmetry when changing $\omega_i \leftrightarrow -\omega_i$, for $i=1,2$. Last but not least, the zero-frequency noise in equilibrium has to vanish.

The vertex function possesses simple poles when $U>2\Gamma$. This is due to the fact that we are using the cutoff-free wide flat-band model. The numerical evaluation
of the noise according to the prescription of Eq. (\ref{eq:RPAfullNPFT}) therefore needs regularization, which is achieved by imposing a hard cutoff scheme which models more realistic finite bandwidths.

 \begin{figure}[!ht]
  \centering
  \includegraphics{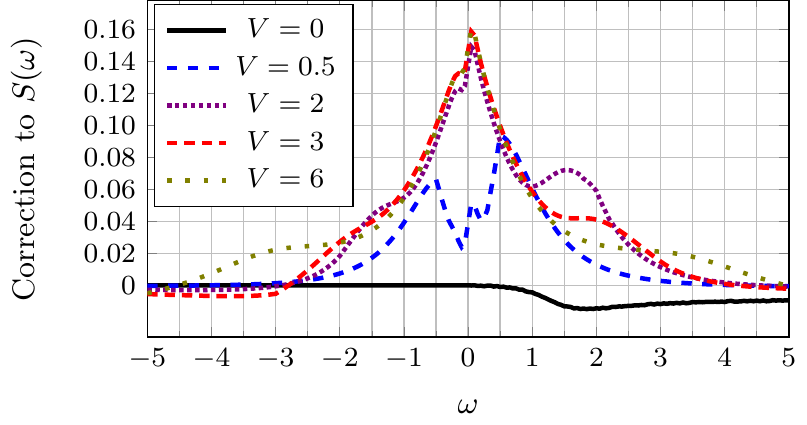}
  \caption{Interaction correction to the noise. The interaction strength is set to $U=0.99$ where $\Gamma^{klmn}$ is regular. The $V=0$ curve is
    enlarged by a factor of 10.}
  \label{fg:RPALowUNoise}
 \end{figure}
We consider the two different regimes $U <> 2\Gamma$ separately. In the $U < 2\Gamma$ case where $\Gamma^{klmn}$ is completely regular, we obtain the results shown in Fig. \ref{fg:RPALowUNoise}
for the interaction induced correction to the non-interacting noise spectrum. In equilibrium, finite interaction causes a small suppression of the noise similar to the result of the
second order perturbation theory. For larger voltages we observe an enhancement in the domain $|\omega|<V$.

 \begin{figure}[!ht]
  \centering
  \includegraphics{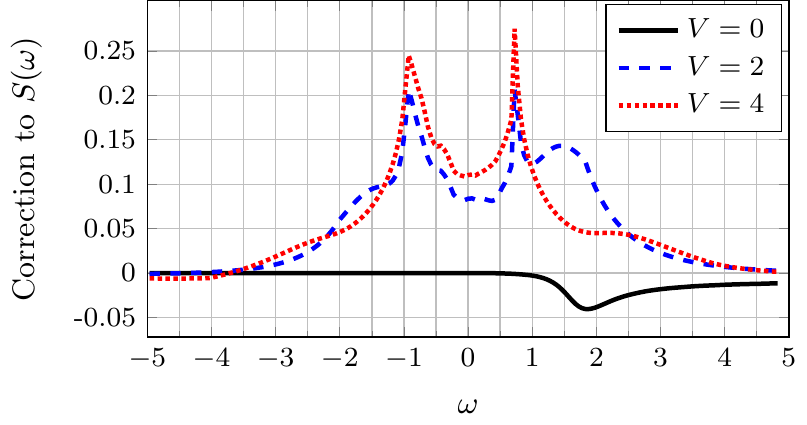}
  \caption{Interaction correction to the noise with $U=1.5$ in the non-regular regime of $\Gamma^{klmn}$. The $V=0$ curve is enlarged by a factor of 10.}
  \label{fg:RPAHighUNoiseOnlyCorrection}
 \end{figure}

More interesting is the case of strong interactions $U>2\Gamma$. As mentioned before, $\Gamma^{klmn}$ develops simple poles in this regime that are also found in the integrand
of Eq.~(\ref{eq:RPAfullNPFT}) as long as $\omega<V/2$.  Due to these poles the correction to the non-interacting noise (Fig. \ref{fg:RPAHighUNoiseOnlyCorrection}) develops sharp peaks at positions  $\pm\omega_{\rm res}<V/2$
that do not seem to be affected by $V$. This turns out to be a generic feature and the distance between the peaks is
solely defined by the interaction strength. The peaks disappear for too small voltages though, see Fig.~\ref{fg:RPAHighUNoiseOnlyCorrection}.

\begin{figure}[!ht]
  \centering
  \includegraphics{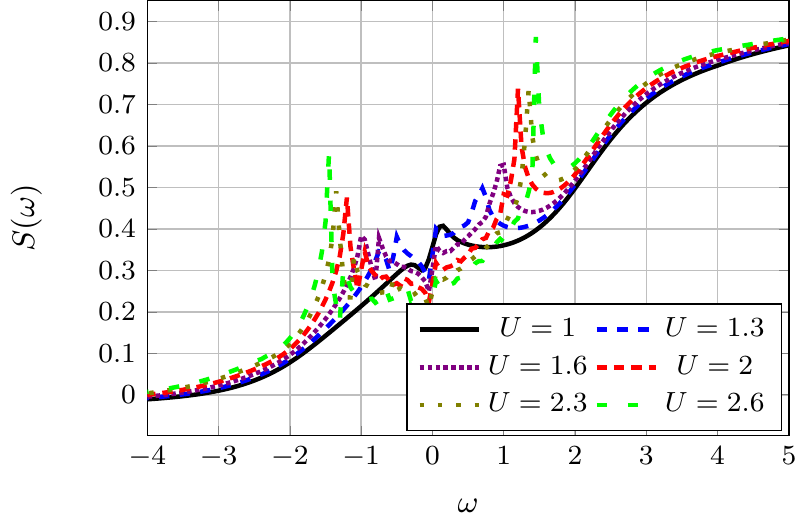}
  \caption{The full noise power for $V=4$ and different interaction strengths calculated with the diagrammatic resummation scheme. For $U=0$ one recovers the
    step-structure.}
  \label{fg:RPAHighUNoise}
 \end{figure}

\begin{figure}[!ht]
  \centering
  \includegraphics{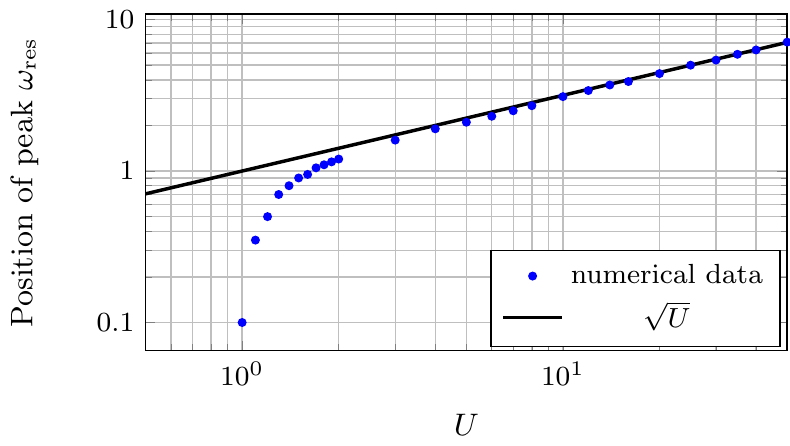}
 \caption{Position of the peaks in the noise spectrum for $V=20$ and various values of $U$. The black line shows a $\sqrt{U}$ behaviour.}
 \label{fg:RPAPeakPos}
\end{figure}

Figure \ref{fg:RPAHighUNoise} shows the full noise spectrum in the non-equilibrium configuration for different Coulomb interaction
strengths. With growing $U$ two distinct peaks at frequencies $\pm \omega_\text{res}$ develop from the non-interacting step-like graph which are not seen in equilibrium
(compare to Fig. \ref{fg:RPAHighUNoiseOnlyCorrection}).  Again, the two peaks correspond to emission and absorption processes and
the noise spectrum is known to exhibit similar peaks in other systems, see for example Ref.\ [\onlinecite{Deblock11072003}].
The position of the peaks $\omega_\text{res}$ as a function of $U$ is plotted in Fig.\ \ref{fg:RPAPeakPos}.
Interestingly, the resulting behavior is highly non-analytic and follows a square-root law with high precision.
This fact points towards the generation of a new energy scale (in a non-perturbative way because otherwise it would be algebraic in $U$), which appears to play a fundamental
role for the dynamical properties of the system. The only plausible option is the Kondo temperature given by\cite{Tsvelick1983}
\begin{equation}
 T_K=\frac{\sqrt{2 U \Gamma}}{\pi} \exp \left(-\pi U/8\Gamma \right) \, .
\end{equation}
The presence of pronounced maxima in the non-equilibrium noise at frequencies $\sim T_K$ can be well-understood. At this energy the system is very sensitive to
fluctuations since the absorbed/emitted energy is comparable to the formation energy of the Kondo singlet,  thus making it an efficient energy dissipation channel.
However, this effect is not directly related to other phenomena of the Anderson model as for example the zero bias conductance peak to which the results of Ref.\ [\onlinecite{PhysRevLett.108.046802}], 
which presents data in the different regime $V\ll U$, are connected.

\section{Summary}
\label{discussion}
In conclusion, we have investigated the frequency-dependent noise in the non-equilibrium Anderson impurity model. Perturbative results in second order for small interaction strength $U$ show
that the noise is suppressed at positive frequencies in equilibrium, compared to the non-interacting setup. In the non-equilibrium situation, however, we observe an increase in the
noise for frequencies related to the applied voltage. Using a diagrammatic resummation scheme we could gain knowledge in the parameter regime which is non-perturbative both
in $\gamma$ and $U$. This approximation shows again a suppression of noise in equilibrium but sharp peaks for finite $V$ at positions that are related to the Kondo
temperature instead of the voltage. This is a consequence of the Kondo temperature defining the energy scale for internal fluctuations of the system.

\acknowledgements

The authors would like to thank  A.~Lebedev, G.~Blatter, G.~Lesovik and and L.~M\"uhlbacher for many interesting discussions. AK is supported by the DFG Grant No.
KO-2235/3-1, CQD and `Enable fund' of the University of Heidelberg.

\bibliography{AIM_noise,references}

\end{document}